\documentclass[10pt,aps,preprint,pra,twocolumn]{revtex4}

\usepackage{graphicx}

\usepackage{multirow}

\begin{document}

\title{Materials properties of out-of-plane heterostructures of MoS$_2$-WSe$_2$ and WS$_2$-MoSe$_2$}
\author{Bin Amin}
\affiliation{Department of Physics, Hazara University, Mansehra, Karakoram Hwy, Dhodial 21120, Pakistan}
\author{Thaneshwor P. Kaloni} 
\affiliation{Department of Chemistry, University of Manitoba, Winnipeg, Manitoba R3T 2N2, Canada}
\author{Georg Schreckenbach}
\affiliation{Department of Chemistry, University of Manitoba, Winnipeg, Manitoba R3T 2N2, Canada}
\author{Michael S. Freund}
\affiliation{Department of Chemistry, University of Manitoba, Winnipeg, Manitoba R3T 2N2, Canada}

\begin{abstract}
Based on first-principles calculations, the materials properties (structural, electronic, vibrational, and optical properties) of out-of-plane heterostructures formed from the transition metal dichalcogenides, specifically MoS$_2$-WSe$_2$ and WS$_2$-MoSe$_2$ were investigated. The heterostructures of MoS$_2$-WSe$_2$ and WS$_2$-MoSe$_2$ are found to be direct and indirect band gap semiconductors, respectively. However, a direct band gap in the WS$_2$-MoSe$_2$ heterostructure can be achieved by applying compressive strain. Furthermore, the excitonic peaks in both monolayer and bilayer heterostructures are calculated to understand the optical behavior of these systems. The suppression of the optical spectrum with respect to the corresponding monolayers is due to interlayer charge transfer. The stability of the systems under study is confirmed by performing phonon spectrum calculations.  
\end{abstract}

\maketitle
Transition metal dichalcogenides (TMDCs) are promising materials for various applications \cite{Zhao,Late,Li1,Chhowalla,Xu}, for example field-effect transistors. Quantum confinement in these semiconductors going from bulk to monolayer results in the transition from indirect to direct band gap semiconductor, which makes them superior in nano-electronics applications as compared to the well studied material graphene that has no band gap.\cite{Thanasis12,He14, santosh1,deep,santosh3,hu,santosh2,mak,santosh4} In addition, due to their distinct electronic properties, TMDC monolayers have been utilized in logic circuits \cite{Radisavljevic11} as well as in memory devices.\cite{Simone13}

Stacking of TMDCs monolayers in order to form heterostructures with van der Waals interactions enables the creation of atomically sharp interfaces \cite{Hong14}, and also provides a route to a wide variety of semiconductor heterojunctions with interesting properties.\cite{Yongji14} Type-II band alignment in these heterostructures has been demonstrated, which reduces the overlap between the electron and hole wave functions by slowing down the charge recombination, which in fact is expected to be an efficient way for light detection and harvesting.\cite{Hong14,Riya} Recently, the long-lived interlayer excitons have been investigated in the MoSe$_2$-WSe$_2$ heterostructure, which indeed opens the large avenue for light-emitting diodes and pholvoltaic devices.\cite{Rivera15}. The structural, electronic, photocatalytic, and optical properties of out-of-plane and in-plane heterostructures of TMDCs have been investigated by employing first-principles calculations.\cite{Amin15} It has been demonstrated that all the out-of-plane heterostructures show an indirect band gap with type-II band alignment. However, a direct band gap can be obtained by the application of tensile strain in specific cases, such as the heterostructures of MoSe$_2$-WSe$_2$ and MoTe$_2$-WTe$_2$. It has also been predicted by theoretical calculations that the direct band gaped bilayer can be generated by alteration of the individual monolayers of TMDCs in the heterostructures.\cite{Humberto13} 

A vertical heterostructure of MoS$_2$-WSe$_2$ has been fabricated by stacking of MoS$_2$ and WSe$_2$ monolayers.\cite{Chiu14} Based on the shifts observed in the Raman and photo-luminescence spectra, it has been found that thermal interactions enhance the coupling between the two monolayers. An atomically thin \textit{p-n} junction diode has been synthesized from a stacked MoS$_2$-WSe$_2$ heterojunction. It shows an excellent current rectification and rapid photo-response with a high quantum efficiency.\cite{Cheng14} Lee et al. \cite{Lee14} also fabricated atomically thin \textit{p-n} heterojunctions from MoS$_2$ and WSe$_2$ monolayers and demonstrated that the tunneling-assisted coupling between MoS$_2$ and WSe$_2$ layers is responsible for the tunability of the electronic and optoelectronic responses. The van der Waals heterostructures of MoS$_2$-WSe$_2$ and MoS$_2$-MoSe$_2$ have been demonstrated using ultra-low frequency Raman spectroscopy.\cite{Chun} It has been shown that a special Raman feature arises from the layer-breathing mode vibration between two incommensurate monolayers. Due to the charge transfer across the interface of the MoS$_2$ and WSe$_2$ heterostructures, a photovoltaic effect has been observed.\cite{Furchi15} Furthermore, a high gate coupling efficiency of about 80\% has been demonstrated for tuning the band offsets at the MoS$_2$-WSe$_2$ vertical interface in dual-gate device.\cite{Tania15}

However, the structural and electronic properties of such heterostructures are not well understood. Therefore, in the present work, comprehensive insight is gained into the physical properties of the MoS$_2$-WSe$_2$ and WS$_2$-MoSe$_2$ heterostructures that are investigated with/without biaxial compressive strain using density functional theory calculations (see supplementary material).\cite{supp} The reduction of the optical spectrum with respect to the corresponding monolayers is analyzed. Furthermore, the stability of the systems under consideration is confirmed by means of the phonon spectrum.


\begin{table}[t]
\begin{tabular}{|c|c|c|c|c|c|c|c|c}
\hline
 & MoS$_2$-WSe$_2$ & WS$_2$-MoSe$_2$ & MoS$_2$-WSe$_2$ &WS$_2$-MoSe$_2$ \\
\cline{1-5} \hline
& \multicolumn{2}{|c|}{\multirow{1}{*}{PBE}} & \multicolumn{2}{|c|}{\multirow{1}{*}{HSE06}} \\
\cline{1-5}
a (\AA) &3.25 &3.25 &3.22&3.22\\
\hline
$d_{spacing}$&3.37 &3.36&3.33&3.34\\
\hline
X--S (\AA) &2.43 &2.43 &2.40&2.40\\
\hline
X--Se (\AA) &2.53 &2.52 &2.50&2.50\\
\hline
$E_b$ (eV) &-0.20&-0.21 &-0.24&-0.25\\
\hline
$E_g$ (eV) &0.33 &0.87 &0.94&1.58\\
\hline
$\Delta_{\rm VB}$ (eV) & 0.46 & 0.19 &0.44&0.18 \\
\hline
$\Delta_{\rm CB}$ (eV) & 0.003 & 0.019 &0.026&0.035 \\
\hline
\cline{1-5}
\end{tabular}
\caption{Lattice constant, interlayer spacing, bond lengths (X=Mo, W), binding energy, band gap, and band splitting (valence and conduction bands) for the out-of-plane heterostructures.}
\end{table}
\begin{figure}[h]
\includegraphics[width=0.17\textwidth,clip]{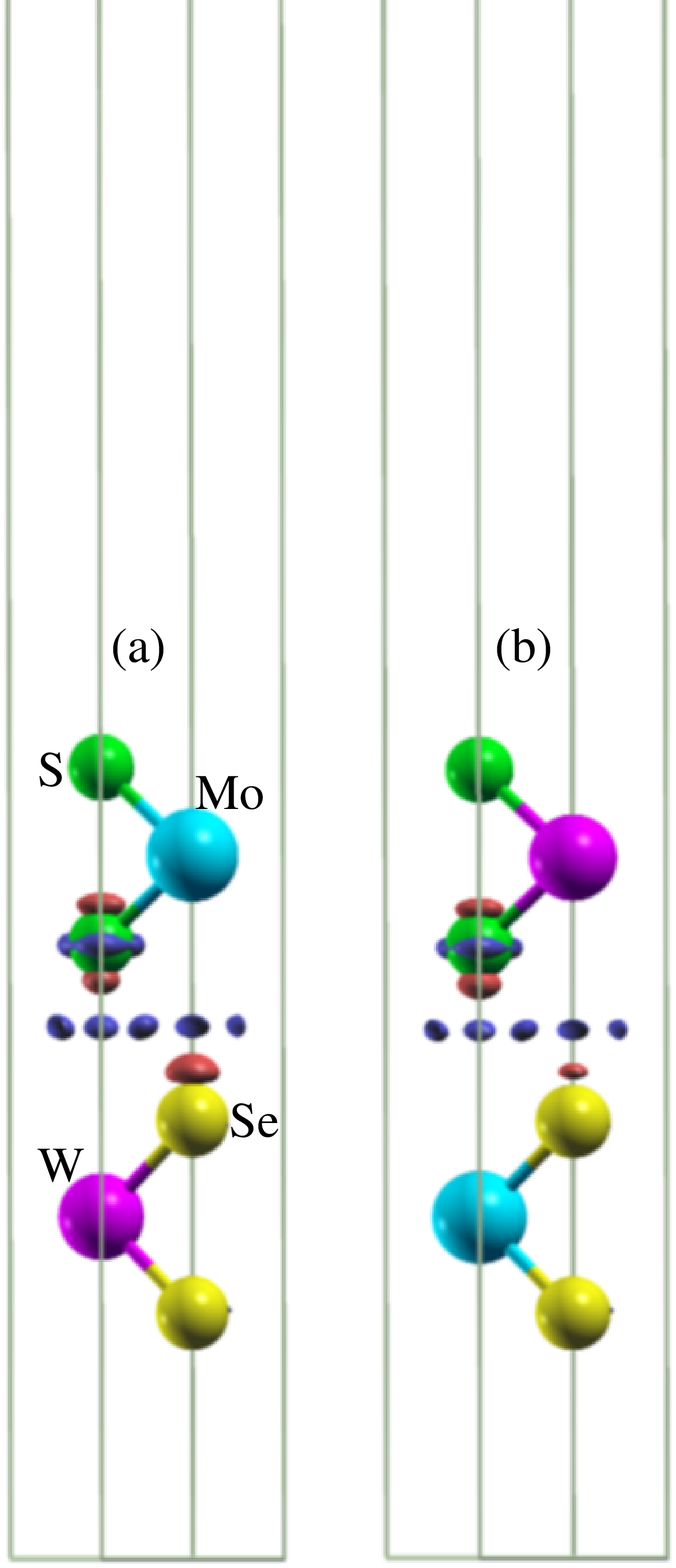}
\caption{Charge density difference isosurfaces for (a) MoS$_2$-WSe$_2$ and (b) WS$_2$-MoSe$_2$ heterostructures with an isovalue of 0.001 eV/\AA, where red and blue color represent positive and negative charges, respectively.}
\end{figure}

The out-of-plane heterostructures were created by using the average of the experimental lattice constant of the constituent monolayers and by keeping one layer on top of the other such that the S/Se atoms of one layer sit on top of the W/Mo of the other layer \cite{Humberto13}, see Fig.\ 1. We have also cross-checked the heterostructures by using the lattice parameters of the constituent monolayers individually. After the relaxation of both the lattice constant and positions, the lattice parameter optimized to the average value of the corresponding monolayers, as summarized in Table I. In principle, the PBE calculations overestimate, while HSE06 calculation give the correct value to the average of the experimental lattice constant of individual monolayers in corresponding heterostructures.\cite{apl} Hence, after relaxation the MoSe$_2$ and WSe$_2$ monolayers suffered from 1.8\% compressive strain, while the MoS$_2$ and WS$_2$ monolayers suffered from the same amount of tensile strain in agreement with Ref.\ \cite{Chiu14}. The obtained value of the interlayer spacing agrees well with experimentally and theoretically obtained values for heterostructures made of MoS$_2$ and WS$_2$ and graphene and WS$_2$ or WSe$_2$.\cite{kaloni-apl,ncom} The calculated bond lengths presented in Table I for both MoS$_2$-WS$_2$ and WS$_2$-MoSe$_2$ agree well with those of the corresponding monolayers (Ref.\ \cite{Kang13}) inferring weak van der Waals interactions between monolayers in the heterostructures. 

The binding energies were obtained as $E = E_{\rm MoX_{2}\textrm{-}WX_{2}} - E_{\rm MoX_{2}} - E_{\rm WX_{2}}$, where $E_{\rm MoX_{2}\textrm{-}WX_{2}}$ (X=S, Se) is the total energy of the heterostructure, and $E_{\rm MoX_{2}}$ and $E_{\rm WX_{2}}$ are the total energies of the corresponding monolayers. The calculated binding energies presented in Table I show that the out-of-plane heterostructures have negative binding energies, which confirms that these systems are energetically favorable. 

\begin{figure}[h]
\includegraphics[width=0.350\textwidth,clip]{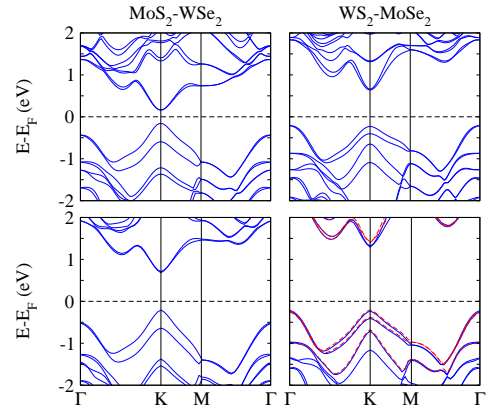}
\caption{Band structures of the out-of-plane heterostructures obtained using PBE (top row) and HSE06 (bottom row) functionals, where red bands are obtained under strain.}
\end{figure}

The band structures presented in Fig.\ 2 show that the heterostructure MoS$_2$-WSe$_2$ is a direct band gap semiconductor, while WS$_2$-MoSe$_2$ is found to be an indirect band gap semiconductor.\cite{Humberto13} 
Normally, the PBE functional underestimates the band gap, therefore, the HSE06 functional is used since it has already been demonstrated that the HSE06 functional provides correct value of the band gap in terms of experimental observation.\cite{apl} The band structures shown in Fig.\ 2 indicate the direct and indirect band gap nature of the heterostructures of MoS$_2$-WSe$_2$ and WS$_2$-MoSe$_2$, respectively. The indirect band gap nature of the WS$_2$-MoSe$_2$ heterostructure is due to the fact that the valence band maximum (VBM) is contributed by the Mo atom, for which the spin-orbit splitting is smaller because of lower atomic number as compared to the W atom. Thus, the VBM lies at the $\Gamma$-point of the Brillouin zone.

To understand the contributions from different atomic sites, we have investigated the weighted band structure of both the heterostructures of MoS$_2$-WSe$_2$ and WS$_2$-MoSe$_2$. It is clear from Fig.\ 3 that the conduction 
band minimum (CBM) at the K-point is due to the Mo $d_{3z^2-r^2}$ orbital and the VBM at the K-point is due to 
the W $d_{x^2-y^2}$ and W $d_{xy}$ orbitals of the MoS$_2$-WSe$_2$ heterostructure. This localization of the VBM and CBM in different monolayers physically separates the electron hole pairs. The strong coupling between the S/Se $p$ orbitals and Mo/W $d_{xz}$ and $d_{yz}$ orbitals leads to a large splitting between their bonding and anti-bonding states, hence these orbitals do not contribute to the band edges. These results show that the VBM and CBM are contributed from different monolayers. This type of arrangement where the holes and electrons are attributed to different layers is known as type-II band alignment.\cite{Hong14} The homogeneous bilayers of TMDCs do not possess this net charge separation, so that an external electric field is required to achieve a type-II band alignment \cite{Ashwin11}. However, there is no electric field in our calculations for the heterostructures that nevertheless show type-II band alignment, which may be due to the intrinsic electric field that arises due to the buckled structures.\cite{k1,k2,k3}

\begin{figure}[t]
\includegraphics[width=0.350\textwidth,clip]{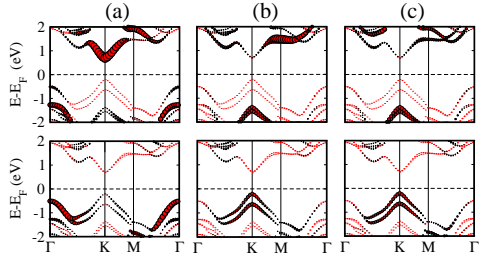}
\caption{Weighted bands of the MoS$_2$-WSe$_2$ out-of-plane heterostructure; Mo (first row), W (second row) for (a) $d_{3z^{2}-r^{2}}$, (b) $d_{x^{2}-y^{2}}$, and (c) $d_{xy}$ orbitals.}
\end{figure}

In contrast to the MoS$_2$-WSe$_2$ heterostructure, the CBM at the K-point of the WS$_2$-MoSe$_2$ heterostructure is due to the W $d_{3z^2-r^2}$ orbital. Without applying strain, the VBM is at the $\Gamma$-point, and it is also due to the W $d_{3z^2-r^2}$ orbital with a major contribution from the Mo $d_{3z^2-r^2}$ orbital. Hence, the heterostructure of WS$_2$-MoSe$_2$ possesses a type-I band alignment, in agreement with the prediction made by Ref. \cite{Hannu13}. Strain engineering is a widely used strategy to achieve tunable band gaps for two-dimensional materials. \cite{Amin14} Therefore, with the application of 0.5\% compressive strain, the VB at the K-point shifts to higher energy, which is due to the Mo $d_{x^2-y^2}$ and Mo $d_{xy}$ orbitals. Whereas, the VBM at the $\Gamma$-point shifts to lower energies. Hence, both the VBM and CBM separate the electron hole pairs to different layers and change the material to the type-II band alignment, see Fig.\ 4. Due to their type-II band alignment, these heterostructures can be used as active materials for fabrication of heterojunction photovoltaic devices; type-II band alignment is usually required for charge separation or formation of $p-n$ junctions.\cite{Lee14}

\begin{figure}[t]
\includegraphics[width=.350\textwidth,clip]{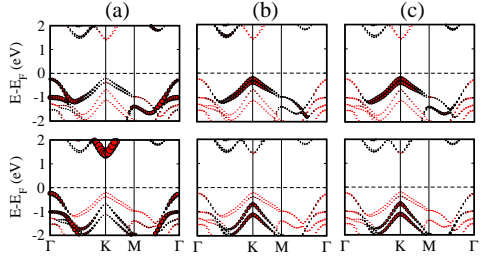}
\caption{Weighted bands of WS$_2$-MoSe$_2$ out-of-plane heterostructure; Mo (first row), W (second row) for (a) $d_{3z^{2}-r^{2}}$, (b) $d_{x^{2}-y^{2}}$, and (c) $d_{xy}$ orbitals.}
\end{figure}

The strong spin-orbit coupling results in a significant valence band splitting ($\Delta_{\rm VB}$) with a 
minute conduction band splitting ($\Delta_{\rm CB}$), see Table 1. The fact is that W contributes much more to VB than Mo, and hence the VB is split. The situation is reversed for the CB, resulting in a split that is very small only, see Fig.\ 3. The valence band splitting in the heterostructure of MoS$_2$-WSe$_2$ was found to be larger than that of pristine MoS$_2$ monolayers ($\Delta_{\rm so}^{PBE}$ $=$0.15 eV and $\Delta_{\rm so}^{HSE}$=0.20 eV) \cite{Kang13} and smaller than that of the pristine WSe$_2$ monolayers ($\Delta_{\rm so}^{PBE}$=0.47 eV and $\Delta_{\rm so}^{HSE}$=0.63 eV) \cite{Kang13}. The same behavior is found for the heterostructure of WS$_2$-MoSe$_2$. The splitting of the valence band (holes) and conduction band (electrons) indicates that the formation of the heterostructure is a promising route to engineer the band gap and splitting. These systems may therefore be interesting for optoelectronic and spintronic devices.\cite{Ochoa13}

\begin{figure}[t]
\includegraphics[width=0.25\textwidth,clip]{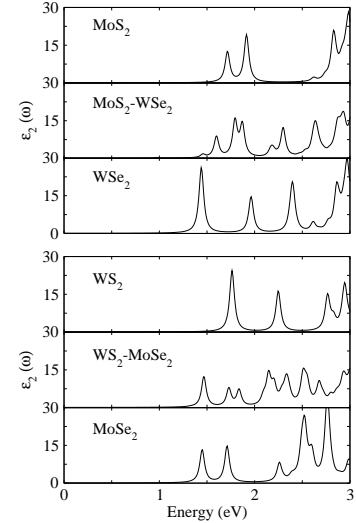}
\caption{Imaginary part of the dielectric function of the heterostructures in comparison with the parent monolayers as a function of photon energy.}
\end{figure}

Further, the imaginary parts of the dielectric function $\epsilon_{2}(\omega)$ of the monolayers and the heterostructures were calculated by solving the Bethe-Salpeter equation. The results presented in Fig.\ 5 show that the optical transitions are dominated by excitons. For monolayer MoS$_2$, excitonic peaks are observed at 1.77 eV and 1.92 eV, while for monolayer WSe$_2$ they appear at 1.50 eV and 1.96 eV. Similarly, these peaks appear at 1.77 eV and 2.24 eV for WS$_2$, while at 1.48 eV and 1.72 eV for monolayer MoSe$_2$. The position of the first excitonic peaks are in good agreement with the experimental values of MoS$_2$ (1.85 eV), WSe$_2$ (1.59 eV), WS$_2$, and MoSe$_2$ (1.56 eV).\cite{Chun, Rivera15,Ashwin12} Fig.\ 5 shows a systematic red shift of the excitonic peaks as the chalcogen atom becomes heavier. In general, for the red shift the band gap decreases, while for the blue shift the band gap increases \cite{Trodahl}. We obtain exciton binding energies of 1.05 eV for MoS$_2$, 0.95 eV for WSe$_2$, 1.04 eV for WS$_2$, and 0.95 eV for MoSe$_2$ monolayers. Strong modifications of the excitons in the heterostructures with respect to the parent monolayers were found, see Fig.\ 5. The excitonic peaks are shifted to 1.6 eV and 1.8 eV in the case of the MoS$_2$-WSe$_2$ heterostructure, in agreement with photoluminescence spectroscopy.\cite{Chun, Furchi15} The remarkable decrease in the photoluminescence intensity of the heterostructures with respect to the parent monolayer has also been reported previously \cite{Chun, Lee14} and is due to the efficient interlayer charge transfer. Note that, in the real experimental situation in the presence of dielectric medium, the position of an excitonic peak might change slightly.\cite{acs}

The charge transfer rate is close to the rate of exciton generation. Due to the type-II band alignment, in case of the MoS$_2$-WSe$_2$ heterostructure the photoexcited electrons in the WSe$_2$ layer tend to flow to the MoS$_2$ layer, and the holes in the MoS$_2$ layer tend to flow to the WSe$_2$ layer. The spatial separation of the electrons and holes therefore suppresses the intralayer optical recombination processes and thus the optical spectrum. To investigate the charge transfer between the constituents monolayers of the heterostructures, the charge density difference was addressed (see Fig.\ 1); calculated as $\Delta \rho= \rho_{MoX_{2}WY_{2}}-\rho_{MoX_2}-\rho_{WY_2}$, where $\rho_{MoX_{2}-WY_{2}}$ is the charge density of the heterostructure, $\rho_{MoX_2}$ is the charge density of the isolated MoX$_2$ monolayer and $\rho_{WY_2}$ is the charge density of the isolated WY$_2$ monolayer. From the Bader population analysis we have fiound that the S atom gains 0.008 e, while the Se atom loses 0.012 e in the MoS$_2$-WSe$_2$ heterostructure. Similarly in case of WS$_2$-MoSe$_2$ Se atoms lose 0.009 e, while S atoms gain 0.016 e, where for unstrained system this amount is 0.009 e for Se and 0.005 e for S. These results are in good agreement with the available values reported in Ref. \cite{Xiangying14}, and show that the interlayer bonding of MoS$_2$ and WSe$_2$ should be rather weak and due to long-range van der Waals forces.

\begin{figure}[t]
\includegraphics[width=0.30\textwidth,clip]{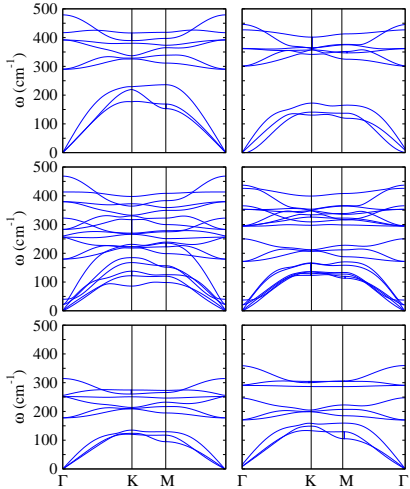}
\caption{Phonon spectrum for MoS$_2$ (top-left), WS$_2$ (top-right), MoS$_2$-WSe$_2$  (middle-left), WS$_2$-MoSe$_2$ (middle-right), WSe$_2$ (bottom-left), and MoSe$_2$(bottom-right). }
\end{figure}

In Fig.\ 6, the calculated phonon spectra for the heterostructures of MoS$_2$-WSe$_2$ and WS$_2$-MoSe$_2$ show good agreement with the experimental data presented in Ref.\ \cite{Chiu14} In the case of the MoS$_2$-WSe$_2$ heterostructure the acoustic and optical vibration modes are at the same energies as for the corresponding monolayers of WSe$_2$ and MoS$_2$. This indicates once more that there are only very small van der Waals interactions between the two layers. The phonon spectra for pristine monolayers of both MoS$_2$ and WSe$_2$ were also calculated for comparison. The characteristic bands observed at 410 cm$^{-1}$ and 380 cm$^{-1}$ are in a good agreement indeed with the experimental values of 403 cm$^{-1}$ and 385 cm$^{-1}$ \cite{Chiu14}, which are due to the A1 mode (out-of-plane vibrations) and E1 mode (in-plane vibrations), respectively, of the MoS$_2$ monolayer in the heterostructure. The bands at about 250 cm$^{-1}$ to 260 cm$^{-1}$ are due to the E1 and A1 modes of the monolayer of WSe$_2$. These bands are also in good agreement with the experimental values of E1 (248 cm$^{-1}$) and A1 (260 cm$^{-1}$).\cite{Chiu14} The positions of these bands are also in agreement with those of the corresponding monolayers with a small red shift (shift to lower energy) with respect to isolated MoS$_2$ and blue shift (shift to higher energy)  with respect to isolated WSe$_2$.\cite{Chiu14} Both of these blue and red shifts with respect to the corresponding monolayers are due to the fact that tensile strain is induced in MoS$_2$ and compressive strain is induced in WSe$_2$ while constructing the heterostructures. The characteristic peaks are well reproduced with respect to the experiments. Although these peaks are not exactly the same as those obtained in experiment, the simulation results provide a good approximation. The positions of the bands in the heterostructure of MoS$_2$-WSe$_2$ with respect to the corresponding monolayers show that the layers are decoupled having a very small van der Waals interaction.

In summary, we have investigated the structural and electronic properties and the vibrational and optical spectra of the van der Waals heterostructures of MoS$_2$-WSe$_2$ and WS$_2$-MoSe$_2$. It was found that MoS$_2$-WSe$_2$ is a direct band gap semiconductor with type-II band alignment. In contrast to the MoS$_2$-WSe$_2$ heterostructure, the WS$_2$-MoSe$_2$ heterostructure is an indirect band gap semiconductor with type-I band alignment. A compressive strain is an efficient way to change the heterostructure of WS$_2$-MoSe$_2$ from indirect to direct band gap nature with type-II band alignment. The splitting of the valence band (holes) and conduction band (electrons) indicates that the formation of the heterostructure is a promising route to engineer the band gap and band splitting, such that these systems would be interesting for optoelectronic and spintronic devices. The reduction of the optical spectrum with respect to the corresponding monolayers is analyzed and the stability of the systems under consideration is confirmed with the help of the phonon spectrum.

GS acknowledges funding from the Natural Sciences and Engineering Council of Canada (NSERC, Discovery
Grant). MSF acknowledges support by NSERC, the Canada Research Chair program, Canada Foundation for Innovation (CFI), the Manitoba Research and Innovation Fund, and the University o f Manitoba.


\begin{thebibliography}{32}
\bibitem{Zhao}Y. Zhao, Y. Zhang, Z. Yang, Y. Yan, and K. Sun, Sci. Technol. Adv. Mater. {\bf 14}, 043501 (2013).

\bibitem{Late}D. J. Late, Y.-K. Huang, B. Liu, J. Acharya, S. N. Shirodkar, J. Luo, A. Yan, D. Charles, U. V. Waghmare, V. P. Dravid \textit{et al.}, ACS Nano {\bf 7}, 4879 (2013).

\bibitem{Li1}Y. Li, H. Wang, L. Xie, Y. Liang, G. Hong, and H. Dai, J. Am. Chem. Soc. {\bf 133}, 7296 (2013).

\bibitem{Chhowalla}M. Chhowalla, H. S. Shin, G. Eda, L.-J. Li, K. P. Loh, and H. Zhang, Nat. Chem. {\bf 5}, 263 (2013).

\bibitem{Xu}K. Xu, Z. Wang, X. Du, M. Safdar, C. Jiang, and J. He, Nanotechnology {\bf 24}, 465705 (2013).

\bibitem{Thanasis12} T. Georgiou, R. Jalil, B. D. Belle, L. Britnell, R. V. Gorbachev, S. V. Morozov, Y. Kim, A. Gholinia, S. J. Haigh, O. Makarovsky \textit{et al.}, Nat. Nanotechnol. \textbf{8}, 100 (2012).

\bibitem{He14} J. He, K. Hummer, and C. Franchini, Phys. Rev. B \textbf{89}, 075409 (2014).

\bibitem{santosh1} P. Zhao, D. Kiriya, A. Azcatl, C. Zhang, M. Tosun, Y.-S. Liu, M. Hettick, J. S. Kang, S. McDonnell, S. KC, \textit{et al.}, ACS Nano \textbf{8}, 10808 (2014)

\bibitem{deep}D. Jariwala, V. K. Sangwan, L. J. Lauhon, T. J. Marks, and M. C. Hersam, ACS Nano \textbf{8}, 1102 (2014).

\bibitem{santosh3}M. Amani, D. Lien, D. Kiriya, J. Xiao, A. Azcatl, J. Noh, S. R. Madhvapathy, R. Addou, S. KC, M. Dubey \textit{et al.}, Science \textbf{350}, 1065 (2015).

\bibitem{hu} Q. H. Wang,	K. Kalantar-Zadeh, A. Kis, J. N. Coleman, and M. S. Strano, Nature Nanotechnol. \textbf{7}, 699 (2012).

\bibitem{santosh2}S. KC, R. C. Longo, R. Addou, R. M. Wallace, K. Cho, Nanotechnology \textbf{25}, 375703 (2014) 

\bibitem{mak}K. F. Mak, C. Lee, J. Hone, J. Shan, and T. F. Heinz, Phys. Rev. Lett. \textbf{105}, 136805 (2010)

\bibitem{santosh4}S. KC, C. Zhang, S. Hong, R. M. Wallace, K. Cho, 2D Materials \textbf{2}, 035019 (2014). 

\bibitem{Radisavljevic11} B. Radisavljevic, M. B. Whitwick, and A. Kis, ACS Nano \textbf{5}, 9934 (2011). 

\bibitem{Simone13} S. Bertolazzi, D. Krasnozhon, and A. Kis, ACS Nano \textbf{7}, 3246 (2013).

\bibitem{Hong14} X. Hong, J. Kim, S. F. Shi, Y. Zhang, C. Jin, Y. Sun, S. Tongay, J. Wu, Y. Zhang, and F. Wang, Nat. Nanotechnol. \textbf{9}, 682 (2014).

\bibitem{Yongji14}  Y. Gong, J. Lin, X. Wang, G. Shi, S. Lei, Z. Lin, X. Zou, G. Ye, R. Vajtai, B. I. Yakobson \textit{et al.}, Nat. Mater. \textbf{13}, 1135 (2014).

\bibitem{Riya}  R. Bose, G. Manna, S. Janaa, and N. Pradhan, Chem. Commun. \textbf{50}, 3074 (2014).

\bibitem{Rivera15} P. Rivera, J. R. Schaibley, A. M. Jones, J. S. Ross, S. Wu, G. Aivazian, P. Klement, N. J. Ghimire, J. Yan, D. G. Mandrus \textit{et al.}, Nat. Commun. \textbf{6}, 6242 (2015).

\bibitem{Amin15} B. Amin, N. Singh, U. Schwingenschl\"ogl, Phys. Rev. B \textbf{92}, 075439 (2015).

\bibitem{Humberto13} H. Terrones, F. Lopez-Urias, and M. Terrones, Sci. Rep. \textbf{3}, 1549 (2013).

\bibitem{Chiu14}  M.-H. Chiu, M.-Y. Li, W. Zhang,  W.-T. Hsu, W.-H. Chang, M. Terrones, H. Terrones, and L.-Jong Li, ACS Nano \textbf{8}, 9649 (2014).

\bibitem{Cheng14} R. Cheng, D. Li, H. Zhou, C. Wang, A. Yin, S. Jiang, Y. Liu, Y. Chen, Y. Huang, and X. Duan, Nano Lett. \textbf{14}, 5590 (2014).

\bibitem{Lee14} C.-Ho Lee, G.-H. Lee, A. M. van der Zande, W. Chen, Y. Li, M. Han, X. Cui, G. Arefe, C. Nuckolls, T. F. Heinz \textit{et al.},  Nat. Nanotechnol. \textbf{9}, 676 (2014).

\bibitem{Chun}  C. H. Lui, Z. Ye, C. Ji, K.-C. Chiu, C.-T. Chou, T. I. Andersen, C. M-. Shively, H. Anderson, J.-M. Wu, T. Kidd \textit{et al.}, Phys. Rev. B \textbf{91}, 165403 (2015).

\bibitem{Furchi15} M. M. Furchi, A. Pospischil, F. Libisch, J. Burgdorfer, and T. Mueller, Nano Lett. \textbf{14}, 4785 (2014).

\bibitem{Tania15} T. Roy, M. Tosun, X. Cao,  H. Fang,  D.-H. Lien, P. Zhao, Y.-Z. Chen, Y.-L. Chueh, J. Guo, and A. Javey, ACS Nano \textbf{9}, 2071 (2015).

\bibitem{supp} See the supplementary material at http://dx.doi.org/.. for details about the computational methods used.






\bibitem{apl}Y. Zhao and D. G. Truhlar, J. Chem. Phys. \textbf{130}, 074103 (2009).

\bibitem{kaloni-apl} T. P. Kaloni, L. Kou, T. Frauenheim, and U. Schwingenschl\"ogl, Appl. Phys. Lett. \textbf{105}, 233112 (2014).

\bibitem{ncom}M.-H. Chiu, C. Zhang, H.-W. Shiu, C.-P. Chuu, C.-H. Chen, C.-Y. S. Chang, C.-H. Chen, M.-Y. Chou, C.-K. Shih, and L.-J. Li, Nat. Commun. \textbf{6}, 7666 (2015). 

\bibitem{Kang13} J. Kang, S. Tongay, J. Zhou, J. Li, and J. Wu, Appl. Phys. Lett. \textbf{102}, 012111 (2013).

\bibitem{Chung13} C.-H. Chang, X. Fan, S.-H. Lin, and J.-L Kuo, Phys. Rev. B \textbf{88}, 195420 (2013).

\bibitem{Ashwin11} A. Ramasubramaniam, D. Naveh, and E. Towe, Phys. Rev. B \textbf{84}, 205325 (2011).

\bibitem{k1}T. P. Kaloni, G. Schreckenbach, M. S. Freund, J. Phys. Chem. C \textbf{118}, 23361 (2014).

\bibitem{k2} T. P. Kaloni, G. Schreckenbach, M. S. Freund, J. Phys. Chem. C \textbf{119}, 3979 (2015).

\bibitem{k3} T. P. Kaloni, M. Modarresi, M. Tahir, M. R. Roknabadi, G. Schreckenbach, M. S. Freund, J. Phys. Chem. C \textbf{119}, 11896 (2015).

\bibitem{Hannu13} H. P. Komsa and A. V. Krasheninnikov, Phys. Rev. B {\bf 88}, 085318 (2013).

\bibitem{Amin14} B. Amin, T. P. Kaloni, and U. Schwingenschl\"ogl, RSC Adv. \textbf{4}, 34561 (2014).

\bibitem{Ochoa13} H. Ochoa and R. Roldan,  Phys. Rev. B \textbf{87}, 245421 (2013).

\bibitem{Ashwin12} A. Ramasubramaniam Phys. Rev. B \textbf{86}, 115409 (2012).

\bibitem{Trodahl}H. J. Trodahl, A. R. H. Preston, J. Zhong, and B. J. Ruck, Phys. Rev. B \textbf{76}, 085211 (2007).

\bibitem{acs} Z. Li, Y. Xiao, Y. Gong, Z. Wang, Y. Kang, S. Zu, P. M. Ajayan, P. Nordlander, and Z. Fang, ACS Nano \textbf{9}, 10158 (2015).

\bibitem{Xiangying14} X. Su, R. Zhang, C. Guo, M. Guo, and Z. Ren, Phys. Chem. Chem. Phys. \textbf{16}, 1393 (2014).

\end{thebibliography}
\end{document}